\documentclass[3p]{elsarticle}

 \pdfoutput=1

\usepackage{hyperref}
\usepackage[usenames,dvipsnames]{color}
\usepackage{amsmath,amssymb}

\journal{Journal of Theoretical Biology}

\usepackage[table,xcdraw]{xcolor}

\begin{document}

\begin{frontmatter}

\title{Modeling of reaction-diffusion transport into a core-shell geometry}

\author[wsumme]{Clarence C. King}
\author[wsumme]{Amelia Ann Brown}
\author[wsumme]{Irmak Sargin}
\author[isumse,isucbe]{K. M. Bratlie}
\author[wsumme]{S. P. Beckman\corref{cor}}
\ead{scott.beckman@wsu.edu}

\cortext[cor]{Corresponding author}
\address[wsumme]{School of Mechanical and Materials Engineering, Washington State University, Pullman, WA 99164, USA}
\address[isumse]{Department of Materials Science and Engineering, Iowa State University, Ames, IA 50014, USA}
\address[isucbe]{Department of Chemical and Biological Engineering, Iowa State University, Ames, IA 50014, USA}

\begin{abstract}
Fickian diffusion into a core-shell geometry is modeled. 
The interior core mimics pancreatic Langerhan islets and the exterior shell acts as inert protection. 
The consumption of oxygen diffusing into the cells is approximated using Michaelis-Menten kinetics. 
The problem is transformed to dimensionless units and solved numerically. 
Two regimes are identified, one that is diffusion limited and the other consumption limited. 
A regression is fit that describes the concentration at the center of the cells as a function of the relevant physical parameters. 
It is determined that, in a cell culture environment, the cells will remain viable as long as the islet has a radius of around 142~$\mu$m or less and the encapsulating shell has a radius of less than approximately 283~$\mu$m.
When the islet is on the order of 100~$\mu$m it is possible for the cells to remain viable in environments with as little as $4.6\times10^{-2}$~mol/m$^{-3}$ O$_2$.
These results indicate such an encapsulation scheme may be used to prepare artificial pancreas to treat diabetes.
\end{abstract}

\end{frontmatter}


\section{Introduction}

Type 1 diabetes is an autoimmune disease that destroys a patient's pancreatic, insulin-producing cells.
Treatment requires blood sugar monitoring and the injection of insulin to keep the patient's blood sugar at safe levels. 
In the worst situations, continual monitoring is necessary via an external insulin pump that is worn under the patient's clothing. 
These extreme situations warrant the development of artificial pancreas technologies to assist stabilizing blood sugar levels.

\begin{figure}[htb]
\centering
\includegraphics[clip, width=\columnwidth]{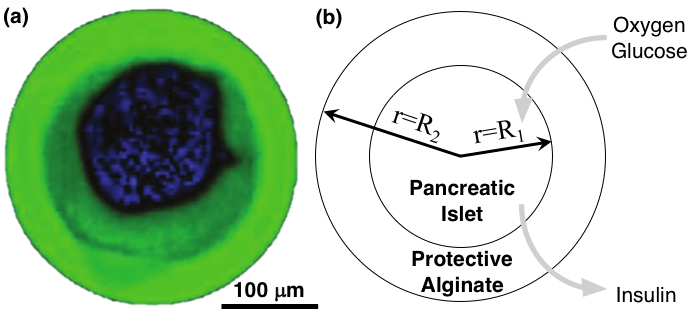}
\caption{Cells encapsulated in an alginate shell, (a) experimentally prepared, reproduced with permission from Ref.~\cite{Ma2013}, and (b) the mathematical model studied here.}
\label{myfig1}
\end{figure}

Transplanting pancreatic islets has been investigated as a possible approach, but rejection is a significant problem~\cite{Ma2013,Bratlie2012,Bygd2016,Bygd2016a}. 
Immunosuppressive drugs  reduce the possibility of rejection, although they are cumbersome to manage and potentially dangerous. 
A better approach is to develop a means to encapsulate pancreatic islets allowing oxygen (16 Da) and glucose (180 Da) to enter and insulin (5.7 kDa) to egress while preventing the body's immune response, \textit{i.e.}, tumor necrosis factor alpha (25.6 kDa) from accessing the cells. 
Such an encapsulation scheme, shown in  Fig.~\ref{myfig1}(a), is viable due to the relatively large size of TNA-$\alpha$ compared to the species being transported~\cite{Opara2013}.
It has been proposed that alginate would be a good choice for the encapsulation layer~\cite{Ma2013}. 
In this manuscript a mathematical model is presented to examine the inward diffusion of molecules, in particular oxygen, through the encapsulation layer and into the cells. 

In 1978 McElwain~\cite{McElwain1978}, building from the 1976 work of Lin~\cite{Lin1976}, modeled the transportation of molecules into a cluster of cells. 
Fickian diffusion was assumed and the consumption of diffusing species was approximated using Michaelis-Menten kinetics. 
As a result of the non-linear consumption term, an analytical solution was not possible. 
A broad range of numerical methods have been used to address variations of this problem including 
the shooting method~\cite{Lin1976}, 
the random choice method~\cite{Sod1986}, 
singular Cauchy methods~\cite{Lima2010}, 
and series decompositions~\cite{Simpson2012}. 
In these, a membrane was sometimes considered present, but its impact on the diffusion rate was not~\cite{Sod1986}. 
Complex modeling has been achieved by finite element methods~\cite{Buchwald2009, Buchwald2011, Buchwald2015, Suszynski2016, Dulong2002a, Dulong2002, Bassom1997} to allow the study of complex geometries, fluid dynamics, and the presence of anoxic cores and cell death within islets. 
These methods are powerful allowing even alternative geometries, such as hollow fibers~\cite{Dulong2002a,Dulong2002,Bassom1997}, to be studied, but their development is costly, frequently requiring the use of multi-physics modeling software.  

Here we extend the theoretical work of McElwain~\cite{McElwain1978} and Lin~\cite{Lin1976}, 
by modeling Fickian diffusion into a spherical core of cells that is encapsulated by a protective shell. 
The consumption of diffusing species is approximated by Michaelis-Menten kinetics.
The results are presented in dimensionless form to allow broad applicability. 
Steady-state  concentrations are calculated for a range of geometries. 
Two regimes of behavior are identified: one that is diffusion limited and another that is consumption limited.
The concentration in the interior of the cell is determined as a function of the Michaelis-Menten parameters, diffusion coefficients, geometry, and the exterior concentration of diffusion species. 

\section{Model}\label{model}

Our model system, shown in Fig.~\ref{myfig1}(b), is based on  experimentally synthesized systems 
such as those described in Ref.~\cite{Bratlie2012} and shown in Fig.~\ref{myfig1}(a). 
Spherical symmetry allows the radial coordinate, $r$, to be the sole geometric descriptor.  
The islet of cells has radius $R_1$ and the encapsulating shell has radius $R_2$. 
The problem is broken into an interior region, solved over the domain $\left(0<r<R_1\right)$, and an exterior region, solved over the domain $\left(R_1<r<R_2\right)$; 
the variables associated with the interior and exterior regions are identified with the subscripts $i$ and $e$ respectively.

Assuming Fickian diffusion and Michaelis-Menten consumption, the steady-state concentration in the interior, $c_{i}\left(r\right)$, is determined from the equation  
\begin{equation}
D_i\left(\frac{2}{r}\frac{d c_i}{d r}+\frac{d^2 c_i}{d r^2}\right)-\frac{V_m c_i}{K_m+c_i}=0,
\label{FLi}
\end{equation}
where $D$ is the diffusion coefficient and $V_m$ and $K_m$ are the Michaelis-Menten parameters corresponding to the maximum consumption rate and the Michaelis constant.
The exterior steady-state concentration, $c_{e}\left(r\right)$, is determined from the equation 
\begin{equation}
D_e\left(\frac{2}{r}\frac{d c_e}{d r}+\frac{d^2 c_e}{d r^2}\right)=0.
\label{FLe}
\end{equation}
These two second-order ordinary differential equations have four boundary conditions. 
Due to spherical symmetry, there is no flux across the center of the sphere, therefore 
\begin{equation}
\frac{\partial c_i}{\partial r} \bigg|_{r = 0}=0.
\label{iniBC1}
\end{equation}
The exterior is in contact with an infinite, well-stirred reservoir with constant concentration $C_2$, yielding 
\begin{equation}
c_e\left(R_2\right)=C_2.
\label{iniBC2}
\end{equation}
The concentration distribution must be continuous at the interface, therefore 
\begin{equation}
c_i\left(R_1\right)=c_e\left(R_1\right).
\label{iniBC3}
\end{equation}
In steady state the flux must be equal at the interface. 
Using Fick's first law,  
\begin{equation}
D_i \frac{d c_i}{d r}\bigg|_{r = R_1}=D_e \frac{d c_e}{d r}\bigg|_{r = R_1}.
\label{iniBC4}
\end{equation}

This system of equations is transformed to dimensionless form to prepare a general solution. 
The dimensionless radial position is 
\begin{equation}
\rho=\frac{r}{R_2}.
\label{NDr}
\end{equation}
The dimensionless interior and exterior concentrations are
\begin{equation}
\chi_i=\frac{c_i}{C_2}
\label{NDci}
\end{equation}
and
\begin{equation}
\chi_e=\frac{c_e}{C_2}.
\label{NDce}
\end{equation}
The dimensionless Michaelis-Menten parameters are
\begin{equation}
\nu=\frac{V_m R_2^2}{D_i C_2}
\label{NDvm}
\end{equation}
and 
\begin{equation}
\kappa=\frac{K_m}{C_2}.
\label{NDkm}
\end{equation}

The differential Eqs.~\ref{FLi} and \ref{FLe} transform to
\begin{equation}
\frac{2}{\rho}\frac{d \chi_i}{d \rho} + \frac{d^2 \chi_i}{d \rho^2} -\frac{\nu \chi_i}{\kappa+\chi_i} = 0 \quad  \left(0<\rho<\frac{R_1}{R_2}\right)
\label{NDFLi}
\end{equation}
and
\begin{equation}
\frac{2}{\rho}\frac{d \chi_e}{d \rho} + \frac{d^2 \chi_e}{d \rho^2} = 0 \quad \left(\frac{R_1}{R_2}<\rho<1\right).
\label{NDFLe}
\end{equation}
The boundary conditions \ref{iniBC1}, \ref{iniBC2}, \ref{iniBC3}, and \ref{iniBC4} transform to 
\begin{equation}
\frac{d \chi_i}{d \rho} \bigg|_{\rho = 0}=0,
\label{NDBC1}
\end{equation}
\begin{equation}
\chi_e\left(1\right)=1,
\label{NDBC2}
\end{equation}
\begin{equation}
\chi_i\left(\frac{R_1}{R_2}\right)=\chi_e\left(\frac{R_1}{R_2}\right),
\label{NDBC3}
\end{equation}
and 
\begin{equation}
D_i \frac{d \chi_i}{d \rho}\bigg|_{\rho = \frac{R_1}{R_2}}=D_e \frac{d \chi_e}{d \rho}\bigg|_{\rho = \frac{R_1}{R_2}}.
\label{NDBC4}
\end{equation}

The first term in  Eqs.~\ref{NDFLi} and \ref{NDFLe} can be eliminated by the substitution  
\begin{equation}
u_i\left( \rho \right)=\rho \chi_i\left( \rho \right)
\end{equation}
and 
\begin{equation}
u_e\left( \rho \right)=\rho \chi_e\left( \rho \right),  
\end{equation}
which yields  
\begin{equation}
\frac{d^2 u_i}{d \rho^2}-\frac{\nu u_i}{\kappa+u_i \rho^{-1}}=0
\label{ODEi}
\end{equation}
and 
\begin{equation}
\frac{d^2 u_e}{d \rho^2}=0.
\label{ODEe}
\end{equation}
The boundary conditions transform to 
\begin{equation}
u_i\left(0\right)=0,
\label{BC1}
\end{equation}
\begin{equation}
u_e\left(1\right)=1,
\label{BC2}
\end{equation}
\begin{equation}
u_i\left(\frac{R_1}{R_2}\right)=u_e\left(\frac{R_1}{R_2}\right),
\label{BC3}
\end{equation}
and 
\begin{equation}
D_i \frac{d u_i}{d \rho}\bigg|_{\rho = \frac{R_1}{R_2}}=D_e \frac{d u_e}{d \rho}\bigg|_{\rho = \frac{R_1}{R_2}}.
\label{BC4}
\end{equation}


\section{Results}

\begin{figure}[htb]
\centering
\includegraphics[clip, width=\columnwidth]{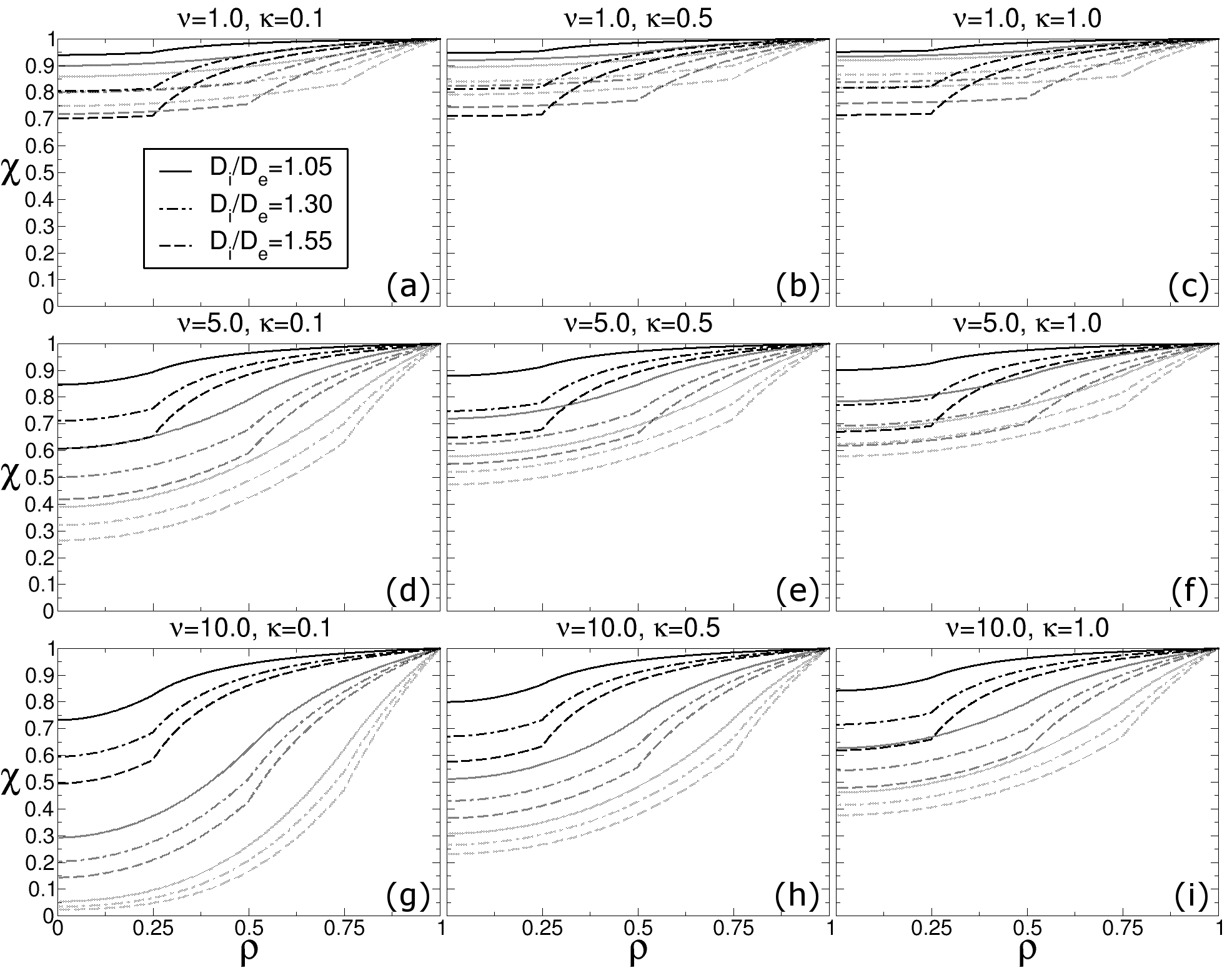}
\caption{
The radial concentration profile as a function of Michaelis-Menten parameters and radius ratio. 
In each of the plots, three radius ratios, $R_1/R_2$, are examined with the values 0.25, 0.50, and 0.75, 
with the darkest line representing 0.25 and the lightest line 0.75. 
Three diffusion coefficient ratios, $D_i/D_e$, are examined with the values 1.05, 1.30, and 1.55. 
The plots in the left, center, and right columns have Michaelis constant, $\kappa$, values 0.1, 0.5, and 1.0 respectively; 
the plots in the top, center, and bottom rows have $\nu$ values 1.0, 5.0, and 10.0 respectively. 
The bottom left plot has the greatest consumption rate and the upper right has the lowest.
\label{myfig2}}
\end{figure}

A numerical solution to differential equations \ref{ODEi} and \ref{ODEe}, 
with boundary conditions \ref{BC1}, \ref{BC2}, \ref{BC3}, and \ref{BC4}, 
is possible by combining the shooting method with the classical fourth-order Runge-Kutta algorithm~\cite{Press1993}. 
The numerical results are verified using the matlab solver \textit{boundary value problem 4c} (\textit{bvp4c}). 
A detailed discussion of the numerical solution is included as supplementary information. 
The dimensionless solutions are presented in Fig.~\ref{myfig2} for a wide range of diffusion coefficient ratios, 
$D_i/D_e$, dimensionless Michaelis-Menten parameters, $\nu$ and $\kappa$, and radius ratio, $R_1/R_2$. 

\begin{figure}[htb]
\centering
\includegraphics[clip, width=\columnwidth]{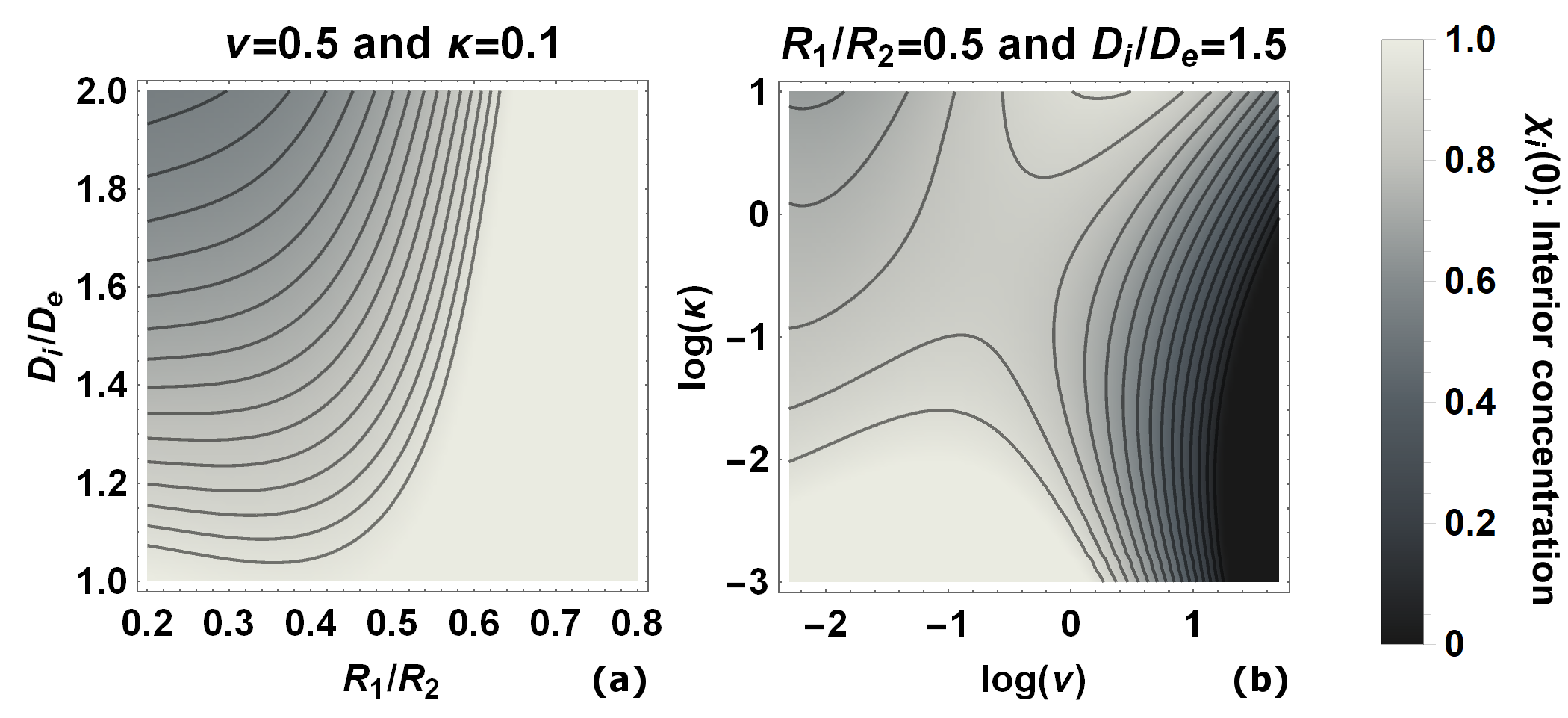}
\caption{
The concentration in the center of the islet, $\chi_i\left(0\right)$, 
(a) calculated as a function of $D_i/D_e$ and  $R_1/R_2$ while holding $\nu=0.5$ and $\kappa=0.1$ constant, and 
(b) calculated as a function of $\log_{10}\left(\kappa\right)$ and  $\log_{10}\left(\nu\right)$ while holding  $R_1/R_2=0.5$ and $D_i/D_e=1.5$ constant.
\label{myfig3}}
\end{figure}

In terms of application, the most important value is the concentration in the center of the islet, $\chi_i\left(\rho=0\right)$.
This concentration is calculated for over $2.7\times10^6$ different combinations of $R_1/R_2$, $D_i/D_e$, $\nu$, and $\kappa$ spanning  parameter space.
A Bayesian-ridge regression is fit to estimate the concentration across parameter space. 
In addition to the initial logarithmic transformations of $\nu$ and $\kappa$, a third-degree polynomial is also applied for the best accuracy of the regression. 
The resulting R$^2$ value is 0.975 and the mean squared error is 0.0344. 
Details of the regression are given in the supplementary information to this manuscript. 
The resulting interior concentration is plotted in Fig.~\ref{myfig3}(a) and (b). 

\section{Discussion}


Fig.~\ref{myfig2}(g) shows the highest consumption rate and \ref{myfig2}(c) the lowest. 
When the consumption rate is low, the concentration profiles are flat inside $r<R_1$.
The interior concentration, $\chi_i\left(0\right)$, is determined by the rate of diffusion through the shell. 
The diffusion coefficient ratios, $D_i/D_e$, separate the solutions some, with the smaller ratios resulting in a higher $\chi_i\left(0\right)$. 
Systems that have thinner encapsulating shells or a higher $D_e$ have a greater $\chi_i\left(0\right)$.
This represents a regime of behavior where diffusion limits the interior solution. 

In contrast the curves separate more when the consumption rate is high. 
In this regime, geometric effects, $R_1/R_2$, become more important than differences in diffusion coefficients,  $D_i/D_e$.
The systems with smaller islets have higher $\chi_i\left(0\right)$ due to there being overall less consumption. 
This represents a regime of behavior where the consumption limits the interior solution. 

It is noteworthy that there is a crossover between these two regimes. 
For example, in Fig.~\ref{myfig2}(a), where $\nu=1.0$, $\kappa=0.1$, and  $D_i/D_e=1.30$, 
the interior solution is approximately equal with $\chi_i\left(0\right)=0.80$ for all three geometries, $R_1/R_2=0.25, 0.50, \mbox{and } 0.75$. 
Increasing the consumption rate shifts the solution toward the consumption limited regime, Fig.~\ref{myfig2}(d), while decreasing the consumption rate shifts the solution toward the diffusion limited regime,  Fig.~\ref{myfig2}(b). 
A similar crossover is observed in Fig.~\ref{myfig3}. 
In Fig.~\ref{myfig3}(a), where the consumption parameters constant, 
the interior concentration is greatest for small values of $D_i/D_e$ and large values of $R_1/R_2$, 
\textit{i.e.}, rapid external diffusion and thin external shells. 
Changing the geometry, $\chi_i\left(0\right)$ begins to drop when $R_1/R_2$ falls below approximately 0.6. 
In Fig.~\ref{myfig3}(b), where  $D_i/D_e$ and $R_1/R_2$ are constant, a significant drop in $\chi_i\left(0\right)$ is observed when $\nu$ becomes larger than approximately 1.75. 
Similar demarcations are observed throughout the dataset described by the regression presented in the  supplementary information. 


The ultimate question being addressed is, \textit{Over what parameter ranges do the cells remain viable?} 
The answer to this can be approximated assuming the onset of hypoxia to be 2.5~mmHg~\cite{Alper1956,Dasu2003}. 
The solubility of O$_2$ in plasma is $1.39\times10^{-3}\mbox{ mM/mmHg}$, but in most tissues 
the solubility is between 1/3 and 1/5 this value~\cite{Carreau2011}.
Therefore, the solubility of O$_2$ in the islet is approximated $1.39\times10^{-3}/4=3.475\times10^{-4}\mbox{ mM/mmHg}$; as a result, the onset of hypoxia is around $8.69\times10^{-4}\mbox{ mol/m}^3$. 
In cell culture experiments the oxygen concentration is approximately 150~mmHg, or 0.21~mol/m$^3$. 
Using this as the surface concentration, $C_2$ in Eq.~\ref{NDci}, the critical dimensionless concentration is $\chi^{\ast}=4.1\times10^{-3}$.  

This value of $\chi^{\ast}$ looks fairly small, but to understand the physical limits the other dimensionless parameters need to be transformed. 
It is clear from Figs.~\ref{myfig2} and \ref{myfig3} that $\nu$, defined in  Eq.~\ref{NDvm}, is highly important for controlling the viability of cells. 
Fig.~\ref{myfig3} demonstrates that when $R_{1}/R_{2}=0.5$ and $D_{i}/D_{e}=1.5$ keeping $\chi_i\left(0\right) > \chi^{\ast}$ requires $\nu < 10.0$. 
Substituting $V_m=3.4\times10^{-2}$~mol/m$^3$s~\cite{Tziampazis1995}, 
$D_{i}=1.3\times10^{-9}$~m$^2$/s~\cite{Evans1981}, 
and $C_2=0.21$ mol/m$^3$,
it is found that keeping $\nu < 10.0$ requires $R_2<283$~$\mu$m, which makes $R_1=142$~$\mu$m.
This is a reasonable size considering that the average Langerhans in humans are between 100 and 150~$\mu$m~\cite{Hellman1959}. 

\begin{figure}[htb]
\centering
\includegraphics[clip, width=\columnwidth]{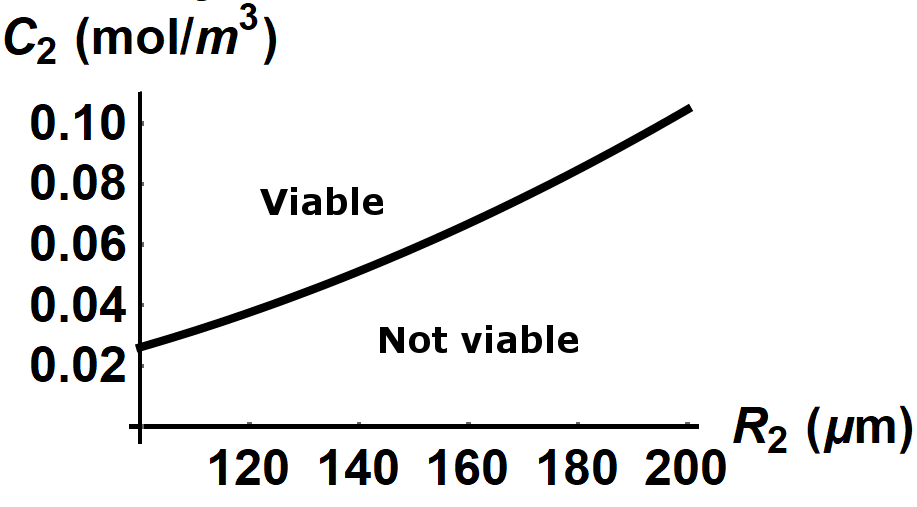}
\caption{
The relationship between $C_2$ and $R_2$ that defines the hypoxia threshold. The region left of the line represents values where the cells are viable, and the right values where hypoxia occurs. 
\label{myfig4}}
\end{figure}

It is likely that \textit{in vivo} the environmental concentration of O$_2$ will be less than 0.21~mol/m$^3$; 
what is the lower limit of $C_2$? 
Assuming that we take the smallest possible islets, around 100~$\mu$m, and have a reasonably thin encapsulation layer, approximately $R_{1}/R_{2}=0.75$, then $R_2=133$~$\mu$m.
Maintaining the limit $\nu<10.0$,  the resulting $C_2$ is  $4.6\times10^{-2}\mbox{ mol/m}^3$, which is 
approximately 25\% the oxygen concentration of cell culture experiments.
Fig.~\ref{myfig4} shows the relationship between $C_2$ and $R_2$ that defines the hypoxia threshold. 
Values to the right of the line results in cell hypoxia. 

%


The calculations here assume uniform porosity in the alginate, cellular tissue, and environment, 
but this is known to not be the case. 
It is possible to modify the model using partition coefficients. 
For the exterior region, the partition coefficient, due to the limited porosity in the cross-linked alginate, is 
\begin{equation}
k_{e}=\frac{c_{e}^{sat}}{C_2},
\label{partcoefe}
\end{equation}
where $c_{e}^{sat}$ is the saturated concentration of the alginate.
Similarly, the partition coefficient for the interior region, due to the limited porosity of tissue, is 
\begin{equation}
k_{i}=\frac{c_{i}^{sat}}{C_2},
\label{partcoefi}
\end{equation}
where $c_{i}^{sat}$ is the saturated concentration of the tissue.

Typically $k_i<k_e<1$, but it is difficult to accurately measure the partition coefficients 
and associated saturated concentrations for alginate and tissue.
The value of the coefficient will vary, depending on the alginate processing, cell density, and hydration. 
As accurate data becomes available, the effect of porosity differences can be included in the model as shown here. 

The differential equations~\ref{ODEi} and \ref{ODEe} are not affected by the partition coefficients, but the boundary conditions are. 
The boundary condition in the center of the sphere, expressed in Eqs.~\ref{iniBC1}, \ref{NDBC1}, and \ref{BC1}, 
is not effected, but the remaining three boundary conditions are. 
The boundary condition at $r=R_2$, Eq.~\ref{iniBC2}, 
incorporates the partition coefficient $k_e$ and transforms to 
\begin{equation}
c_e\left(R_2\right)=k_{e}C_2.
\label{iniBC2part}
\end{equation}
In dimensionless form this becomes 
\begin{equation}
\chi_e\left(1\right)=k_{e},
\label{NDBC2part}
\end{equation}
and further transforms to
\begin{equation}
u_e\left(1\right)=k_{e}.
\label{BC2part}
\end{equation}
The boundary condition ensuring continuity at the core-shell interface, 
Eq.~\ref{iniBC3}, 
incorporates the partition coefficients and transforms to 
\begin{equation}
k_e c_i\left(R_1\right) = k_i c_e\left(R_1\right).
\label{iniBC3part}
\end{equation}
In dimensionless form this becomes 
\begin{equation}
k_e \chi_i\left(\frac{R_1}{R_2}\right) = k_i \chi_e\left(\frac{R_1}{R_2}\right),
\label{NDBC3part}
\end{equation}
and further transforms to
\begin{equation}
k_e u_i\left(\frac{R_1}{R_2}\right) = k_i u_e\left(\frac{R_1}{R_2}\right).
\label{BC3part}
\end{equation}
The boundary condition ensuring the continuity of flux across the core-shell interface, 
Eq.~\ref{iniBC4}, 
incorporates the partition coefficients and transforms to
\begin{equation}
k_e D_i \frac{d c_i}{d r}\bigg|_{r = R_1} = k_i D_e \frac{d c_e}{d r}\bigg|_{r = R_1}.
\label{iniBC4part}
\end{equation}
In dimensionless form this becomes 
\begin{equation}
k_e D_i \frac{d \chi_i}{d \rho}\bigg|_{\rho = \frac{R_1}{R_2}} = k_i D_e \frac{d \chi_e}{d \rho}\bigg|_{\rho = \frac{R_1}{R_2}},
\label{NDBC4part}
\end{equation}
and further transforms to
\begin{equation}
k_e D_i \frac{d u_i}{d \rho}\bigg|_{\rho = \frac{R_1}{R_2}} = k_i D_e \frac{d u_e}{d \rho}\bigg|_{\rho = \frac{R_1}{R_2}}.
\label{BC4part}
\end{equation}

\section{Conclusions}

A dimensionless model of Fickian diffusion into a spherical core-shell geometry, 
where the interior mimics cellular tissue and the exterior inert protective encapsulation, is created. 
The concentration of diffusing species is calculated for a wide range of dimensionless parameters. 
Using these results, along with physical constants from literature, it is concluded that it is possible to devise an encapsulation-based artificial pancreas technology. 

Using physically reasonable values for the Michaelis-Menten parameters, diffusion constants, and environmental oxygen, it is determined that the cells will remain viable only if the islet is less than 142~$\mu$m, which is the size of actual pancreatic islets in humans.
Further, if the islet is held at its minimum size, 100~$\mu$m, it is possible to keep the cells alive if the environmental O$_2$ concentration is approximately 25\% that of cell cultures, around $4.6\times10^{-2}$~mol/m$^3$. 
Throughout these calculations there is uncertainty; for example the diffusion coefficient of O$_2$ in alginate depends strongly on the processing. 
These variations in physical constants will impact the quantitative accuracy of the calculated results, but the overall conclusions will not be impacted. 

The geometry used in the model presented here is clearly an oversimplification. 
Irregularities in islet and alginate shape will lead to regions with shorter and longer diffusing paths.
The consumption rate of oxygen is not truly represented by Michaelis-Menten kinetics because it coupled to the concentration of glucose, which diffuses slower than O$_2$. 
There will likely be interface effects, both at the exterior and where the alginate and cells meet. 
It is possible to develop highly adjustable finite element simulations to include these complexities, and the results presented here provide compelling motivation for the development of such calculations. 

The mathematical model presented here is potentially a starting point for addressing other engineering problems. 
For example, it might find use in studying spherical catalysts in which the exterior has undergone permanent damage. 
These results might also serve as the starting point for a model of tumor growth inside tissues, for example, in soft tissue sarcomas. 

\section*{References}
\bibliography{mybibfile}

\begin{thebibliography}{10}
\expandafter\ifx\csname url\endcsname\relax
  \def\url#1{\texttt{#1}}\fi
\expandafter\ifx\csname urlprefix\endcsname\relax\def\urlprefix{URL }\fi
\expandafter\ifx\csname href\endcsname\relax
  \def\href#1#2{#2} \def\path#1{#1}\fi

\bibitem{Ma2013}
M.~Ma, A.~Chiu, G.~Sahay, J.~C. Doloff, N.~Dholakia, R.~Thakrar, J.~Cohen,
  A.~Vegas, D.~Chen, K.~M. Bratlie, T.~Dang, R.~L. York, J.~Hollister-Lock,
  G.~C. Weir, D.~G. Anderson, {Core-shell hydrogel microcapsules for improved
  islets encapsulation.}, Advanced healthcare materials 2~(5) (2013) 667--72.
\newblock \href {http://dx.doi.org/10.1002/adhm.201200341}
  {\path{doi:10.1002/adhm.201200341}}.

\bibitem{Bratlie2012}
K.~M. Bratlie, R.~L. York, M.~A. Invernale, R.~Langer, D.~G. Anderson,
  {Materials for diabetes therapeutics.}, Advanced healthcare materials 1~(3)
  (2012) 267--84.
\newblock \href {http://dx.doi.org/10.1002/adhm.201200037}
  {\path{doi:10.1002/adhm.201200037}}.

\bibitem{Bygd2016}
H.~Bygd, K.~Bratlie,
  \href{http://www.mdpi.com/2073-4360/8/12/422}{{Investigating the Synergistic
  Effects of Combined Modified Alginates on Macrophage Phenotype}}, Polymers
  8~(12) (2016) 422.
\newblock \href {http://dx.doi.org/10.3390/polym8120422}
  {\path{doi:10.3390/polym8120422}}.
\newline\urlprefix\url{http://www.mdpi.com/2073-4360/8/12/422}

\bibitem{Bygd2016a}
H.~C. Bygd, K.~M. Bratlie, \href{http://doi.wiley.com/10.1002/jbm.a.35700}{{The
  effect of chemically modified alginates on macrophage phenotype and
  biomolecule transport}}, Journal of Biomedical Materials Research Part A
  104~(7) (2016) 1707--1719.
\newblock \href {http://dx.doi.org/10.1002/jbm.a.35700}
  {\path{doi:10.1002/jbm.a.35700}}.
\newline\urlprefix\url{http://doi.wiley.com/10.1002/jbm.a.35700}

\bibitem{Opara2013}
E.~C. Opara, J.~P. McQuilling, A.~C. Farney, {Microencapsulation of pancreatic
  islets for use in a bioartificial pancreas}, Methods in Molecular Biology
  1001 (2013) 261--266.
\newblock \href {http://dx.doi.org/10.1007/978-1-62703-363-3_21}
  {\path{doi:10.1007/978-1-62703-363-3_21}}.

\bibitem{McElwain1978}
D.~L.~S. McElwain, {A re-examination of oxygen diffusion in a spherical cell
  with Michaelis- Menten Oxygen uptake kinetics}, Journal of Theoretical
  Biology 71~(2) (1978) 255--263.

\bibitem{Lin1976}
S.~Lin, {Oxygen diffusion in a spherical cell with nonlinear oxygen uptake
  kinetics}, Journal of Theoretical Biology 60~(2) (1976) 449--457.
\newblock \href {http://dx.doi.org/10.1016/0022-5193(76)90071-0}
  {\path{doi:10.1016/0022-5193(76)90071-0}}.

\bibitem{Sod1986}
G.~A. Sod, {A numerical study of oxygen diffusion in a spherical cell with the
  Michaelis-Menten oxygen uptake kinetics.}, Journal of mathematical biology
  24~(3) (1986) 279--289.
\newblock \href {http://dx.doi.org/10.1007/BF00275638}
  {\path{doi:10.1007/BF00275638}}.

\bibitem{Lima2010}
P.~M. Lima, L.~Morgado, {Numerical modeling of oxygen diffusion in cells with
  Michaelis-Menten uptake kinetics}, Journal of Mathematical Chemistry 48~(1)
  (2010) 145--158.
\newblock \href {http://dx.doi.org/10.1007/s10910-009-9646-x}
  {\path{doi:10.1007/s10910-009-9646-x}}.

\bibitem{Simpson2012}
M.~J. Simpson, A.~J. Ellery, {An analytical solution for diffusion and
  nonlinear uptake of oxygen in a spherical cell}, Applied Mathematical
  Modelling 36~(7) (2012) 3329--3334.
\newblock \href {http://dx.doi.org/10.1016/j.apm.2011.09.071}
  {\path{doi:10.1016/j.apm.2011.09.071}}.

\bibitem{Buchwald2009}
P.~Buchwald, {FEM-based oxygen consumption and cell viability models for
  avascular pancreatic islets.}, Theoretical Biology and Medical Modeling
  6~(5).
\newblock \href {http://dx.doi.org/10.1186/1742-4682-6-5}
  {\path{doi:10.1186/1742-4682-6-5}}.

\bibitem{Buchwald2011}
P.~Buchwald, {A local glucose-and oxygen concentration-based insulin secretion
  model for pancreatic islets}, Theoretical Biology and Medical Modeling
  8~(20).
\newblock \href {http://dx.doi.org/10.1186/1742-4682-8-20}
  {\path{doi:10.1186/1742-4682-8-20}}.

\bibitem{Buchwald2015}
P.~Buchwald, S.~R. Cechin, J.~D. Weaver, C.~L. Stabler, {Experimental
  evaluation and computational modeling of the effects of encapsulation on the
  time-profile of glucose-stimulated insulin release of pancreatic islets},
  Biomedical Engineering Online 14~(28).
\newblock \href {http://dx.doi.org/10.1186/s12938-015-0021-9}
  {\path{doi:10.1186/s12938-015-0021-9}}.

\bibitem{Suszynski2016}
T.~M. Suszynski, E.~{S. Avgoustiniatos}, K.~K. Papas, {Oxygenation of the
  Intraportally Transplanted Pancreatic Islet}, Journal of Diabetes Research
  2016.

\bibitem{Dulong2002a}
J.~Dulong, C.~Legallais, S.~Darquy, G.~Reach, {A novel model of solute
  transport in a hollow-fiber bioartificial pancreas based on a finite element
  method}, Biotechnology and Bioengineering 78~(5) (2002) 576--582.

\bibitem{Dulong2002}
J.~L. Dulong, C.~Legallais, {Contributions of a finite element model for the
  geometric optimization of an implantable bioartificial pancreas}, Artif
  Organs 26~(7) (2002) 583--589.

\bibitem{Bassom1997}
A.~P.~I. Bassom, A.~Ilchmann, H.~Voss, {Oxygen diffusion in tissue preparations
  with Michaelis-Menten kinetics}, J Theor Biol 185~(1) (1997) 119--127.

\bibitem{Press1993}
W.~H. Press, S.~A. Teukolsky, W.~T. Vetterling, B.~P. Flannery, Numerical
  Recipes in FORTRAN; The Art of Scientific Computing, 2nd Edition, Cambridge
  University Press, New York, NY, USA, 1993.

\bibitem{Alper1956}
T.~Alper, H.-F. P., Role of oxygen in modifying the radiosensitivity of e. coli
  b, Nature 178 (1956) 978--979.

\bibitem{Dasu2003}
A.~Dasu, I.~Toma-Dasu, M.~Karlsson, Theoretical simulation of tumour
  oxygenation and results from acute and chronic hypoxia, Physics in Medicine
  and Biology 48 (2003) 2829--2842.

\bibitem{Carreau2011}
A.~Carreau, B.~El~Hafny-Rahbi, A.~Matejuk, C.~Grillon, C.~Kieda, Why is the
  partial oxygen pressure of human tissues a crucial parameter? small molecules
  and hypoxia, J. Cell Mol. Med. 15~(6) (2011) 1239--1253.

\bibitem{Tziampazis1995}
E.~Tziampazis, A.~Sambanis, \href{http://www.ncbi.nlm.nih.gov/pubmed/7766095
  http://doi.wiley.com/10.1021/bp00032a001}{{Tissue Engineering of a
  Bioartificial Pancreas: Modeling the Cell Environment and Device Function}},
  Biotechnology Progress 11~(2) (1995) 115--126.
\newblock \href {http://dx.doi.org/10.1021/bp00032a001}
  {\path{doi:10.1021/bp00032a001}}.
\newline\urlprefix\url{http://www.ncbi.nlm.nih.gov/pubmed/7766095
  http://doi.wiley.com/10.1021/bp00032a001}

\bibitem{Evans1981}
N.~T. Evans, P.~F. Naylor, T.~H. Quinton,
  \href{http://www.ncbi.nlm.nih.gov/pubmed/7280375}{{The diffusion coefficient
  of oxygen in respiring kidney and tumour tissue.}}, Respiration physiology
  43~(3) (1981) 179--88.
\newline\urlprefix\url{http://www.ncbi.nlm.nih.gov/pubmed/7280375}

\bibitem{Hellman1959}
B.~Hellman, Actual distribution of the number andvolume of the islets of
  langerhans indifferent size classes in non-diabetic humans of varying ages.,
  Nature 184 (1959) 1498 -- 1499.

\end{thebibliography}

\end{document}